# Novel Memory Structures in QCA Nano Technology

**Ali H. Majeed*[1,2], Esam Alkaldy[2], Mohd S. Zainal[1] and Danial MD. Nor[1]**
[1]FKEE, UTHM, Johor, Malaysia
[2]Electrical Department, Faculty of Engineering, University of Kufa, Kufa, Iraq

**Correspondence**
*Ali H. Majeed
Electrical Department, Faculty of Engineering,
University of Kufa, Kufa, Iraq
Email: alih.alasady@uokufa.edu.iq

**Abstract**
*Quantum-dot Cellular Automata (QCA) is a new emerging technology for designing electronic circuits in nanoscale. QCA technology comes to overcome the CMOS limitation and to be a good alternative as it can work in ultra-high-speed. QCA brought researchers attention due to many features such as low power consumption, small feature size in addition to high frequency. Designing circuits in QCA technology with minimum costs such as cells count and the area is very important. This paper presents novel structures of D-latch and D-Flip Flop with the lower area and cell count. The proposed Flip-Flop has SET and RESET ability. The proposed latch and Flip-Flop have lower complexity compared with counterparts in terms of cell counts by 32% and 26% respectively. The proposed circuits are designed and simulated in QCADesigner software.*

## I. INTRODUCTION

The limitations in CMOS technology such as high lithography, short channel effects, and power consumption encouraged scientists to think about alternatives. Many nanotechnologies were emerged to overcomes these limitations such as Single Electron Transistor [1, 2], Carbon Nanotube Field-Effect Transistor [3-7], FinFET [8-10] and Quantum-dot Cellular Automata (QCA). QCA technology was introduced for the first time by Lent et al in 1993 [11] and it is reliability was studied in [12]. QCA building block is a square shape that has four dots and two electrons. This technology is reviewed in details by [13, 14] QCA depends on the principle of electron's repulsion [15]. The memory unit is very important in all electronic circuits. Designing efficient memory units in QCA technology still in the competition. The researchers in QCA technology looking for finding a good circuit with minimum complexity (cells and area). This paper introduces a novel latch with minimum complexity then uses this latch for designing an optimum form of D Flip-Flop with set/reset ability.

## II. BACKGROUND

The basic QCA technology block is a quantum cell. This cell consists of four holes (dots) in addition to a couple of electrons. The electrons configuration inside the cell gives it a certain polarization. Only two configurations can be formed by a QCA cell so, it can represent the binary numbers. Cell with polarized -1 represents binary 0 and the binary 1 can be represented by a cell with polarized +1 as shown in Figure 1.

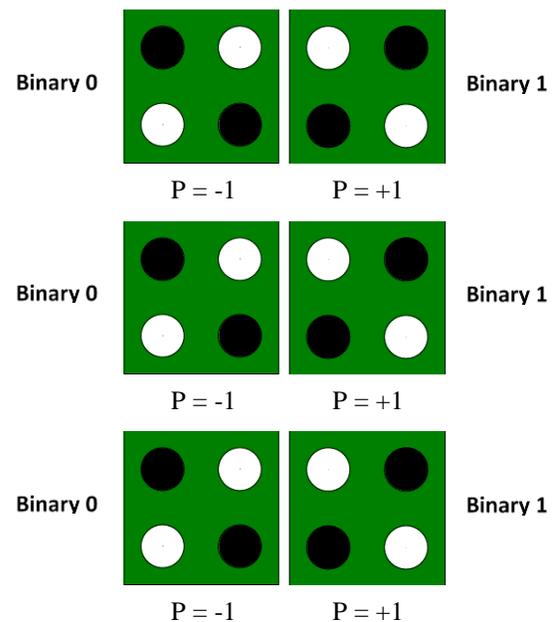

Fig.1: QCA cell configuration

The dominant gate in QCA technology is the majority gate where AND/OR gates can be built using this gate. Majority gate constructed with 5 cells as illustrated in Figure 2. The basic gates (AND/OR) designs shown in Figure 3. Multi-input majority gate are also available as in [16].

In QCA, there is also another important block beside the majority gate called an inverter. The two main structures of the inverter depicted in Figure 4.

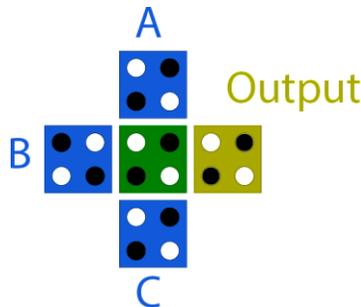

Fig. 2: QCA majority gate

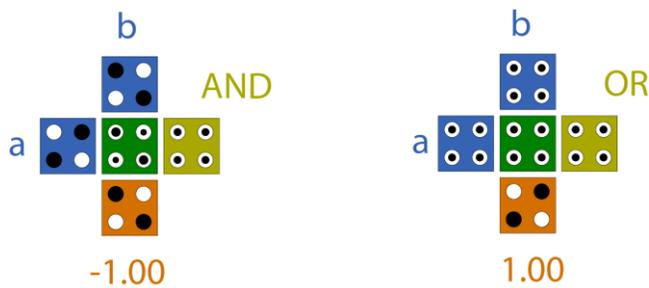

Fig. 3: The general form of AND and OR gates

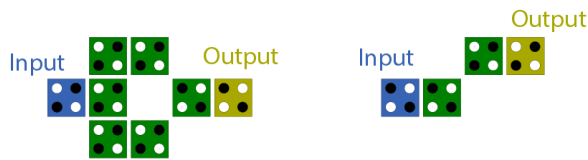

Fig. 4: Two main structures of QCA-inverter

Binary wire in QCA technology can be constructed by a set of cells put beside each other as shown in Figure 5.

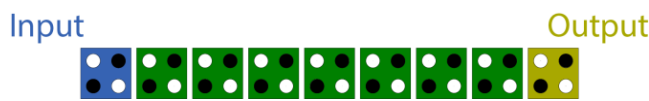

Fig. 5: QCA wire

Generally, QCA circuits need a clock signal for many reasons such as synchronization and control of the direction of flowing signals [17, 18]. The large circuits in QCA can be divided into four zones each zone has four clock phases (switch, hold, release and relax) as shown in Figure 6 to ensure adiabatic switching and keep the circuit close to the ground state. The clock signal controls the barriers between dots to allow or prevent the flowing of electrons. QCA cells give true value during hold state only. Therefore, in this phase, the cells will act as the drive to the adjacent cells.

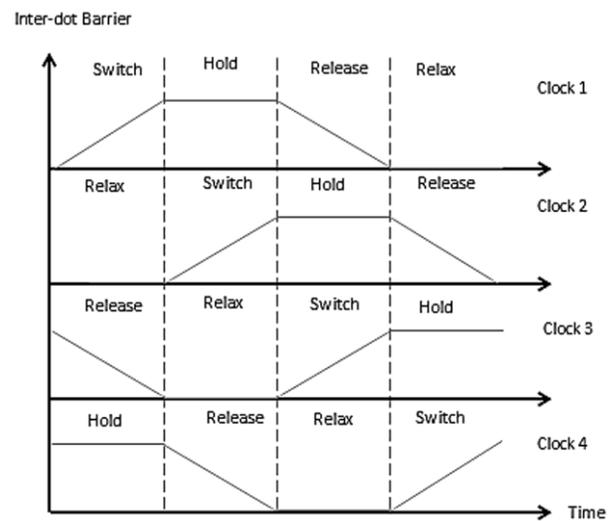

Fig. 6: QCA-clock signal

### III. RELATED WORK

In QCA literature, the counters and flip-flops are discussed a lot [19]. Many structures of D-lathes and flip-flops were introduced in QCA technology. All of them aiming to find an optimum structure in terms of area and cell counts. Some best relevant of such circuits are shown in Figure 7. In Figure 7(a) proposed by [20], the authors proposed D flip-flop without set/reset ability where D flip-flop having the ability to set and reset the output is proposed in [21] as shown in Figure 7(b). In Figure 7(c), the design reported in [22] is shown with better performance.

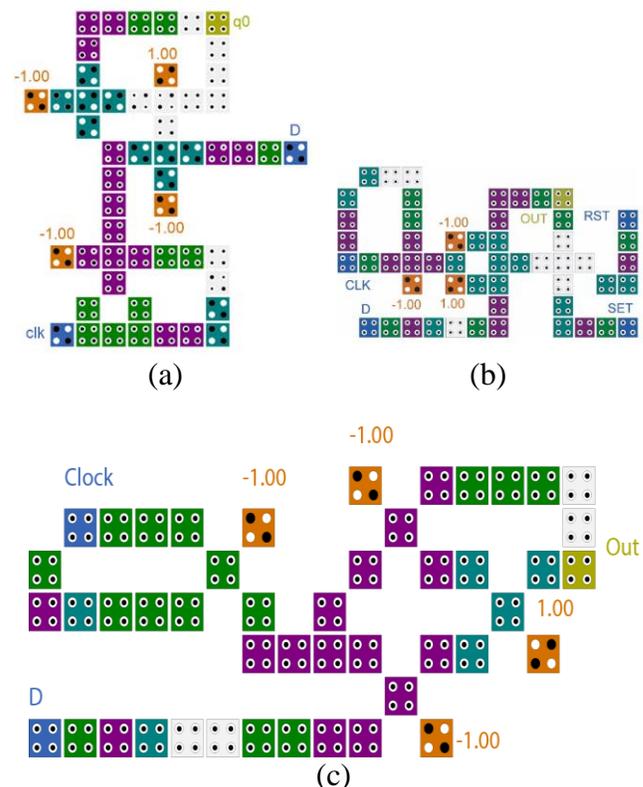

Fig. 9: D flip-flop proposed (a) in [20] (b) in [21] (c) in [22]

## IV. PROPOSED DESIGN

In this section, a novel QCA-design of D latch will be proposed. The block diagram that will be followed to design the latch is depicted in Figure 8. Although many multiplexers are presented in QCA literature discussed in [23], the multiplexer structure that will be utilized in this work is proposed in [24]. So, the proposed D latch is illustrated in Figure 9. From the proposed D latch, D Flip-flop in both positive and negative edge triggers are constructed by adding small circuit called level to edge converter as shown in Figure 10.

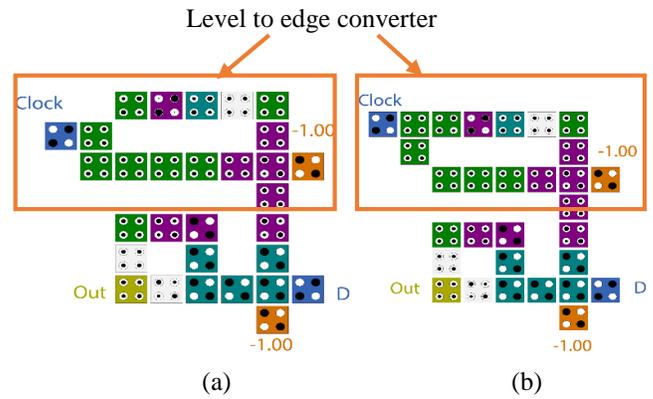

Fig. 10: Proposed D Flip-Flop (a) +Ve edge trigger (b) -Ve edger trigger

From designs shown in Figure 10, it can add set and reset feature by adding majority gate at the output. The proposed Flip-Flop with set/reset ability will be as depicted in Figure 11. The functionality table of the proposed flip-flop is detailed in Table 1. The proposed structures in this work are very efficient compared to the early reported ones in terms of number of cells and layout area, and adopting them to larger circuits will cause significant improvements.

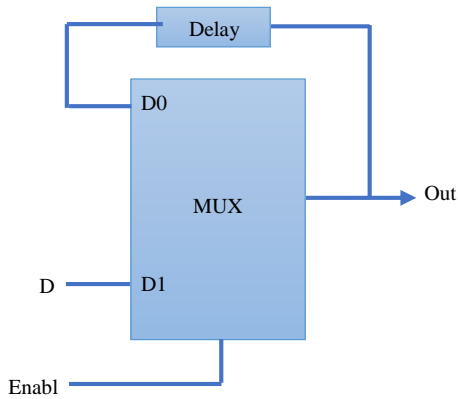

Fig. 8: D flip-flop Block diagram

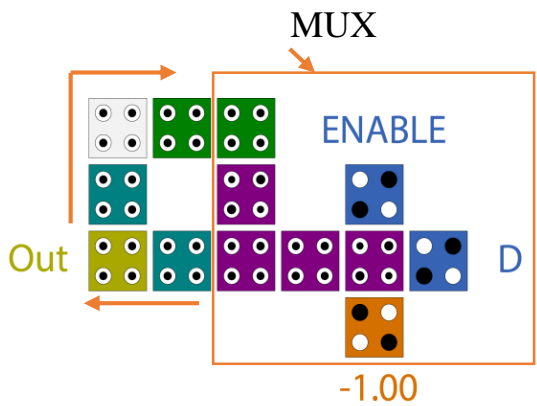

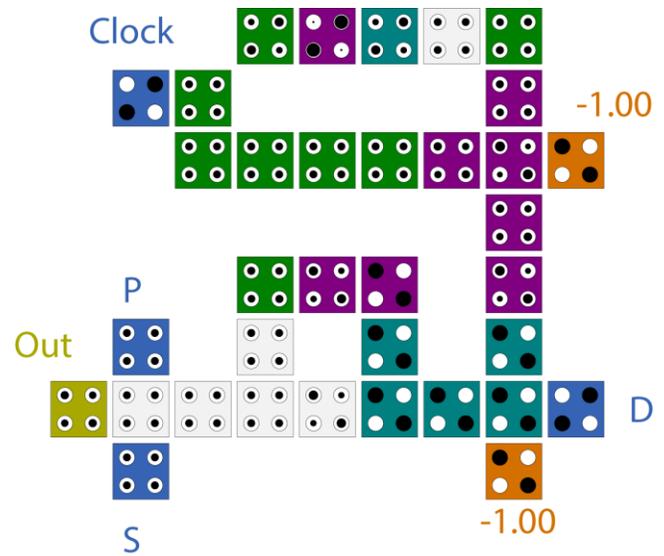

Fig. 11: Proposed D Flip-Flop (+Ve) with Set/reset ability

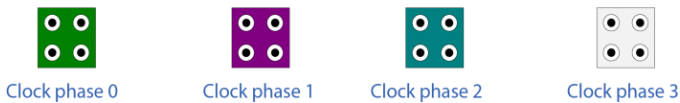

Fig. 9: Proposed D latch

TABLE 1:
Proposed D Flip-Flop (+Ve) functionality table

| P | S | Clock transition | D | Out(t+1) |
|---|---|---|---|---|
| 0 | 0 | x | x | 0 (reset) |
| 1 | 1 | x | x | 1 (set) |
| 0 | 1 | 0 to 1 | 0 | 0 |
| 1 | 0 | | | |
| 0 | 1 | 0 to 1 | 1 | 1 |
| 1 | 0 | | | |
| 0 | 1 | 1 to 0 | X | Out |
| 1 | 0 | | | |

## V. SIMULATION RESULT AND COMPARISON

This section will explain the simulation result of the proposed circuits using QCADesigner software [25]. The output waveforms prove that the proposed circuits are free of error as illustrated in Figure 12. The proposed D latch and D Flip-Flop are compared with counterparts as detailed in Tables 2 and 3. It's clear from Table 2 that the proposed D Latch has a better structure than the best reported in terms of cell count of about 32 %, and in terms of geometry and layout the proposed design is optimized and this improves the circuit area by 50 %. The comparison results in Table 3 for the D-FF are also encouraging. Although the improvement in cell count is around 26 %, the geometry optimization of the proposed structure is very effective and caused circuit area improvement of 57 %. The major cause of the improvements in the proposed structures is the used Multiplexer structure.

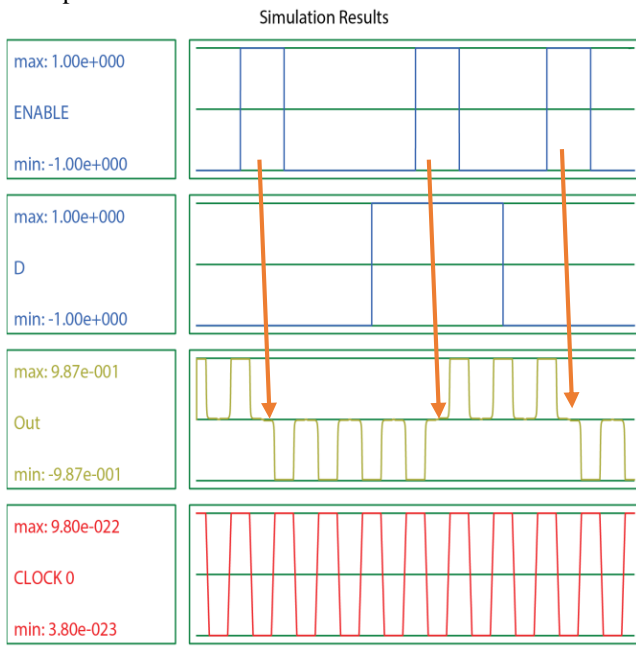

Fig. 12: Output waveforms of the proposed D-latch

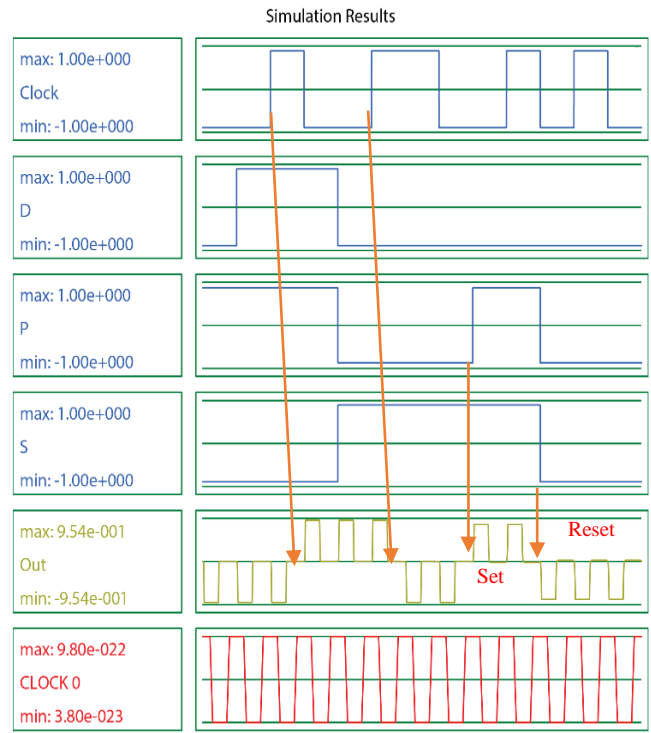

(a)

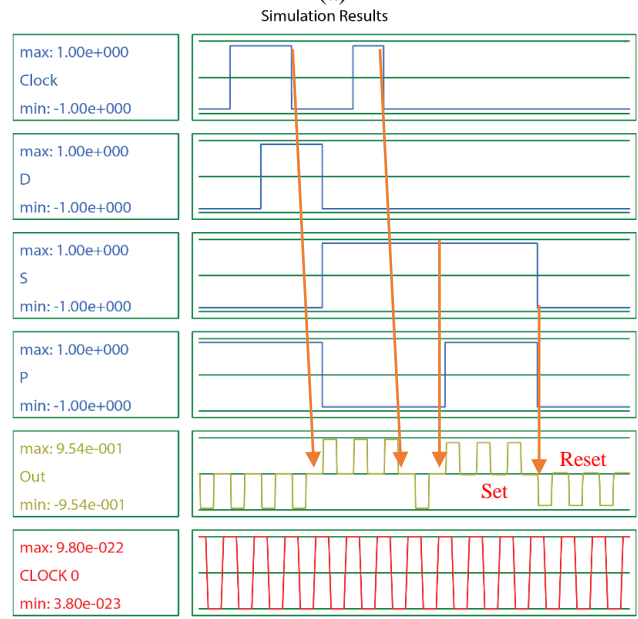

(b)

Fig. 13: Output waveforms of the proposed D-FF with set/reset (a) positive edge (b) negative edge

TABLE 2:
Proposed D latch comparison table

| Design | Area (μm2) | Cell count | Clock phases |
|---|---|---|---|
| D Latch in [26] | 0.05 | 48 | 4 |
| D Latch in [27] | 0.06 | 43 | 4 |
| D Latch in [28] | 0.02 | 28 | 2 |
| D Latch in [21] | 0.02 | 19 | 3 |
| Proposed D latch | 0.01 | 13 | 3 |

TABLE 3:
Proposed D Flip-Flop comparison table

| Design | Area (μm2) | Cell count | Clock phases | S/R ability |
|---|---|---|---|---|
| D-FF in [29] | 0.11 | 84 | 11 | No |
| D-FF in [30] | 0.06 | 56 | 10 | No |
| D-FF in [21] | 0.04 | 53 | 9 | Yes |
| D-FF in [22] | 0.07 | 47 | 7 | No |
| Proposed D-FF | 0.03 | 35 | 8 | Yes |

## VI. Conclusion

This paper presents novel structures of memory units (D latch and D Flip-Flop) in QCA technology. The proposed circuits are in optimal form in terms of area and cell counts. The proposed flip flop has the ability to set the output and reset it. The proposed circuits have more efficient than previously published in almost all metrics as explained in comparison tables. The output waveforms indicate that the proposed structures are free of error.